\documentclass[a4paper,11pt]{article}
\pdfoutput=1 % if your are submitting a pdflatex (i.e. if you have
\usepackage{jcappub} % for details on the use of the package, please
\usepackage[T1]{fontenc} % if needed
\usepackage{amsmath}

\title{\boldmath Can strangelets be detected in a large LAr neutrino detector?}

\author[a]{Mihaela P\^arvu}
\author[a,1]{Ionel Lazanu \note{Corresponding author.}}

\affiliation[a]{University of Bucharest, Faculty of Physics, POBox MG-11 Bucharest-Magurele, Romania}

\emailAdd{mihaela.parvu@unibuc.ro}
\emailAdd{ionel.lazanu@g.unibuc.ro}

\abstract{Predicted as possible bound states of up, down and strange quarks, strangelets could be more energetically favourable and more stable than nuclear matter. In this paper we explore the possibility of detecting such particles with the future large liquid argon detectors developed for neutrino physics. Using signals from ionization and scintillation, as well as measuring the range, we suggest that a calorimetric TPC detector is able to put in evidence and to discriminate between light strangelets and normal isotopes at intermediate energies.}

\begin{document}
\maketitle
\flushbottom

\section{Introduction}

Particles in the Standard Model (SM) have lifetimes spanning an enormous range of magnitudes, from stable particle as electron, through the proton with mean lifetime $> 2.1 \times 10^{29}$ years (or stable), up to the Z boson ($\approx2 \times 10^{-25}$ s). Models beyond the SM (BSM) typically predict new particles with a variety of lifetimes that can generically have lifetimes long compared to SM particles at the weak scale. The reasons include a) approximate symmetries that stabilize the long-lived particle (LLP), b) small couplings between the LLP and lighter states and c) suppressed  phase space available for decays \cite{Alimena}. 

If these particles are sufficiently long-lived, they can exist in the cosmos either as Big Bang relics, are produced in the astrophysical processes, or as secondary collision products and thus must be searches in accelerator-based experiments at higher energies \cite{Alimena} or in non-collier experiments \cite{Burdin:2014xma}.
The discovery of LLPs would address important questions in modern physics: such as the origin and composition of dark matter and the unification of the fundamental forces.

The new generation of neutrino experiments opens a window for searches of physics BSM because: a) the combination of the high intensity proton beam facilities and massive detectors for precision measurements of neutrino oscillation parameters will help make BSM physics reachable even in low energy regimes; b) large-mass detectors with highly precise tracking and energy measurements, excellent timing resolution, and low energy thresholds will enable the searches for BSM phenomena also from cosmogenic origin \cite{Arguelles:2019xgp}. Thus, experiments as DUNE or Hyper-Kamiokande have capabilities to make BSM physics more reachable. In fact, the experiments have included this goal in their scientific program \cite{Abi:2020kei, Abe:2016ero}.

The early Universe started with equal matter and antimatter abundance in thermal equilibrium, unless they have a very weak self-interaction cross section. The particles would therefore be created when other particles annihilate with each other but the rate at which this would occur is limited by the mass of the particle and the temperature of the Universe. The species in question would continue to follow the relic abundance set by this Boltzmann suppression until it goes out of thermal equilibrium, at which point whatever number of particles that are present per unit entropy is ‘‘frozen in’’. At late times, entropy density corresponds to photon density, but if other species freeze out after the species in question, they will dump their entropy into the photon bath and further reduce the relic abundance of the initial massive particle \cite{Burdin:2014xma}. A summary of possible types of particles are: LLPs as heavy leptons and hadrons, WIMPs , Axion, Dark Atoms, Mirror Matter, Strange Quark Matter,  Magnetic monopoles (in different theories: Dirac, as dyson particle, GUT monopole, extra dimensions, string theory), Q-balls, Black Holes.

Continuing the previous work \cite{Lazanu:2020qod}, in the present paper we take LAr as an example of active medium to explore its recording capabilities of signals produced by strangelets.

\section{Strangelets}

\subsection{About the structure and properties of the strangelets}

The normal nuclear matter is composed by up and down quarks even if measurements also show the presence of a strange component in the structure of the proton, for example. Witten \cite{Witten:1984rs} and independently Farhi and Jaffe \cite{Farhi:1984qu}, from theoretical considerations, suggested that at densities slightly above nuclear matter density, a strange quark matter composed of up, down, and strange quarks in roughly equal numbers could exist and it could be stable. This is strange nuclear matter, dominated by strong interaction. In the frame of several models existing in the literature, see for example \cite{Gilson:1993zs, Madsen:2000kb} it is suggested that the Strange Quark Matter (SQM) could be energetically favourable to nuclear matter, depending on the values of relevant parameters. 

General features, common to all types of strangelets consist in: a) masses per baryon below the mass per nucleon in nuclei (by assumption of stability),  and b) a very low charge-to-mass ratio. From observational point of view, it is expected that high A/Z ratio compared to normal nuclei is a feature associated with a possible strange structure. In environment  they could be neutralised by electrons and form unusual atoms or ions. In accord with the model developed by Jes Madsen \cite{Madsen:2004vw}  for the strangelets with colour-flavour locked quark matter, a $Z = 0.3A^{2/3}$ dependence is predicted and for the strangelets without pairing (‘‘ordinary’’ strangelets), dependencies as $Z = 0.1A$ for $A\ll$700 and $Z = 8A^{1/3}$ for $A\gg$700 are predicted, assuming a strange quark mass of 150 MeV.
An important problem is repesented by the mass range of the existence of long-lived strangelets. In the absence of experimental evidence, all the estimations are done in the frame of different models and, furthermore, are model dependent. For example, Alford and co-workers, \cite{Alford:2006bx} address the stability of electrically neutral bulk quark matter with respect to fragmentation into a charge-separated strangelet and  electron crust, and the almost precisely equivalent question of whether large strangelets are stable with respect to fission into smaller strangelets. The estimation of the minimum value of $A$ for long-lived strangelets is model-dependent. A typical minimum $A$ values are in the range $A_{min} \approx  10 \div 600$, with up to $Z \leq 70$ for ordinary strangelets, or $A \approx 680$ with $Z \approx 105$ in the case of-colour–flavour locked strangelets. These predictions are all consistent with the result $Z \approx 0.1 A$ from the literature. If smaller strangelets with mass number smaller than the minimum $A$ are created, they will decay rapidly. If the ground state composition of strangelet consists of equal numbers of quarks of the three types of quark flavours - which is the most favourable state, thus it is neutral. Other, are expected to possess a small positive electric charge can be neutralised by capturing electrons. 
In accord with Madsen \cite{Madsen:2005uv}, the mass of a strangelet as a function of baryon number, in the lowest energy ground state is given as:

\begin{equation}
M(A)=874 \mathrm{[MeV]} \times A + 77 \mathrm{[MeV]} \times A^{2/3} + 232 \mathrm{[MeV]}  \times A^{1/3},
\end{equation}
if the mass of strange quark is supposed to be 150 MeV and a bag constant $B^{1/4}=145$ MeV. In this parameterization bulk quark matter is bound by 56 MeV/baryon and only strangelets with baryon number $A > 23$ are stable relative to ordinary nuclei, but this limit is proportional with both parameters of the model - increasing the strange quark mass and/or the bag constant moves the stability to higher $A$.
However, these assumptions are not confirmed and should be treated with precaution. 

Although the analysis of the production mechanisms of strangelets is not essential for our studies, it is useful to remember that there are several scenarios proposed in the literature; cosmological origin, in energetic nuclear collisions, collisions of strange stars for example. If the cosmologic origin is considered, these particles are remnants of the cosmic QCD phase transition \cite{Witten:1984rs,Alam:1997ij,Bhattacharyya:1999dm,Bhattacharyya:1999jj}. In high energetic nuclear collisions associated the formation of quark-gluon plasma could be possible and thus strangelets \cite{Greiner:1987tg,Greiner:1988pc,Greiner:1991us}. Collisions of strange stars and different other astrophysical processes represent also a possibility to the formation of strangelets. See for example the references \cite{Madsen:1998uh,Biswas:2012cz,Biswas:2014dya}.

\subsection{Direct candidates for strangelets and exotic events interpreted as signals of SQM}

By consulting the literature, a small number of singular experimental events interpreted as SQM can be identified.

\begin{enumerate}
\item 
Events Saito \cite{Saito:1990ju}. In the analysis of background events from a counter-technique experiment to study cosmic-ray nuclei above 8 GV/c rigidity two events were observed with $Z\sim14$, $A\sim350$, and 450 amu, and which cannot be accounted for by more conventional background. It is concluded that as a confirmation of SQM at a flux level of $\sim6\times10^{-9} \mathrm{cm^{-2}} \mathrm{s^{-1}} \mathrm{sr^{-1}}$ and the theoretical analysis was done by Kasuya and co-workers \cite{Kasuya:1991ef} considering an average value $A = 370$ amu and $Z = 14$.

\item
Price’s event \cite{Price:1976mr} with $Z = 46$ and $A>1000$ is fully consistent with SQM considering the ratio $Z/A$. 

\item
An event with $Z = 20$ and $A = 460$ has been reported in \cite{Ichimura:1990ce}.

\item
NA52 hypothetical event. One event with a mass 7.4 GeV/c$^2$ and charge number $Z=-1$, in  the $p/Z=-100$GeV/c rigidity settings was observed \cite{Klingenberg:1996wu}.

\item
AMS-01 events. These events were identified by the Alpha Magnetic Spectrometer (AMS-01) experiment, when the detector was flown on the space shuttle Discovery during flight STS–91 (1998). First candidate has $Z/A = 0.114 \pm 0.01$ and a kinetic energy of 2.1 GeV, and has been assigned to $^{16}$He. The second candidate was reconstructed as having a nuclear charge of $Z = 8$ and a mass $A=54^{+8}_{-6}$ which we will denote as $^{54}$O \cite{Han:2009sj}.

\end{enumerate}

In literature are also mentioned a number of phenomena that are both puzzling and extremely unusual or singular. i) Centauro events for which anomalous composition of secondary particles with almost no neutral pions present observed in emulsion chambers \cite{Lattes:1980wk}. A re-analysis of this case done by Ohsawa and co-workers \cite{Ohsawa:2006mg} claimed an incorrect analysis of the experimental data. ii) anomalous cosmic ray bursts from Cygnus X - 3 \cite{Baym:1985tn,Shaw:1985zr}. iii) extraordinary high luminosity gamma-ray bursts from the supernova remnant N49 in the Large Magellanic Cloud \cite{Alcock:1986xm}.
Rybczynski and Zbigniew Włodarczyk –see for example reference \cite{Rybczynski:2019hys} where all previous calculations are cited, show that the abundance of strangelets in the primary cosmic ray flux is about $2.4 \times 10^{-5}$ at the same energy per particle. The analysis of the EAS data offers a unique possibility to observe possible imprints of strangelets arriving from outer space. These authors tried to explain the muon excess observed by the Pierre Auger Observatory assuming the existence of strangelets. Results suggest that the possible contribution of strangelets is negligible in the limits of the precision of data.

Expected strangelet flux. Monreal \cite{Monreal:2005dg}  derived an equation to estimate the abundance of strangelets  in cosmic-ray propagation model. The author's work assumes a strangelet production rate of 10$^{-10}$ solar masses ($M_\odot$)  per year in our Galaxy, with 10$^{-5}$$M_\odot$  of strange matter released per collision, and a collision every 30000 y. It includes the effects of acceleration, interstellar propagation, and solar modulation using several phenomenological models. If we assume that all of the released strange-star material is converted into strangelets of charge $Z$ and and mass $A$ (GeV/c$^2$ ), the maximum flux $F$ is: 

\begin{equation}
F = 2 \times 10^5 \mathrm{[m^{-2} y^{-1} sr^{-1}]} \times A^{-0.467} Z^{-1.2} (R^{1.2})_{cutoff}.
\end{equation}

$R_{cutoff}$ is the either the geomagnetic cutoff or the solar-modulation cutoff, whichever is greater. Similar estimations of the strangelet flux in cosmic rays was done by Madsen in two successive papers \cite{Madsen:2004vw,Madsen:2005uv}. Madsen has done distinct predictions: 

-for “ordinary” strangelets:
\begin{equation}
F_{total} \approx 2.5 \times 10^5 \mathrm{[m^{-2} y^{-1} sr^{-1}]} A^{-1.067} Z^{-0.6} \Phi^{-0.6}_{500} \Lambda .
\end{equation}

-for "CFL" strangelets:
\begin{equation}
F_{total} \approx 2.8 \times 10^4 \mathrm{[m^{-2} y^{-1} sr^{-1}]}  Z^{-2.2} \Phi^{-0.6}_{500} \Lambda .
\end{equation}

In Table \ref{fluxes} some estimations of the fluxes of strangelets is done, considering “colour flavour locked quark matter and "ordinary” possibilities.

\begin{table}[h]
\begin{tabular}{|c|c|c|c|c|}
\hline
A &  \multicolumn{2}{c|}{Colour flavour locked quark matter} & \multicolumn{2}{c|}{Ordinary}\\
\hline
 & (Z,A) & Flux [m$^{-2}$ yr$^{-1}$ sr$^{-1}$] & (Z,A) & Flux [m$^{-2}$ yr$^{-1}$ sr$^{-1}$]\\
\hline
6 & (1,6) & 28000 & (1,6) & 36953 \\
\hline
14 & (2,14) & 6094 & (2,14) & 9872 \\
\hline
18 & (2,18) & 6094 & (2,18) & 7550 \\
\hline
22 & (2,22) & 6094 & (2,22) & 6095 \\
\hline
36 & (3,36) & 2497 & (4,36) & 2378 \\
\hline
56 & (4,56) & 1326 & (6,56) & 1163 \\
\hline
70 & (5,70) & 812 & (7,70) & 836 \\
\hline
350 & (14,350) & 84 & (35,350) & 57 \\
\hline
450 & (17,450) & 55 & (45,450) & 38 \\
\hline
\end{tabular}
\caption{Strangelets fluxes estimations.}
\label{fluxes}
\end{table}

The strangelets are influenced by the solar wind when entering the inner parts of the Solar System. Here, $\Phi_{500}=\Phi/(500 \mathrm{MeV})$. A good fit to the solar modulation of the cosmic ray spectrum can be given in terms of a potential model, where the charged particle climbs an electrostatic potential of order  $\Phi=500$ MeV (the value changes by a factor of less than 2 during the 11 (22) year solar cycle). The second quantity, $\Lambda$, has the explicit form: 
\begin{equation}
\Lambda = \left(\frac{\beta_{SN}}{0.005} \right)^{1.2} \left(\frac{0.5 \mathrm{cm^{-3}}}{n} \right) \left(\frac{M}{10^{-10}M_\odot \mathrm{yr^{-1}}} \right) \left(\frac{1000 \mathrm{kpc^{3}}}{V} \right) \left(\frac{930 \mathrm{MeV}}{m_0 c^2} \right).
\end{equation}

The minimal rigidity is assumed to be given by the speed of a typical supernova shock wave, $\beta_{SN} \sim 0.005$. The strangelet spectrum after acceleration in supernova shocks is assumed to be a standard power law in rigidity with index 2.2 as derived from observations of ordinary cosmic rays. $n$ - denotes the average hydrogen number density per cubic centimeter ($n \sim 0.5 \mathrm{cm^{-3}}$) when averaging over denser regions in the galactic plane and dilute regions in the magnetic halo). For a total production rate of $M=10^{-10} M_\odot \mathrm{yr}^{-1}$ of baryon number $A$ strangelets spread evenly in an effective galactic volume $V$. A typical value for the effective galactic volume confining cosmic rays is $V=1000$ kpc$^3$.

Integral upper limits on the strangelet flux in cosmic rays was therefore set for particles with charge 1 $\le$ $Z$ $\le$ 8 and mass 4 $\le$ $A$ $\le$ 1.2 $\times$ 10$^{5}$ in terms of baryon number ($A$) was measured by PAMELA \cite{PAMELA:2015lnr}. In figure 3 from the cited article are indicated these limits: $1.5\times{10}^3\div1.5\times{10}^7\left[\ \mathrm{m^{-2}y^{-1}{sr}^{-1}}\right]$, relative constant versus baryon number, a different behaviour of the upper fluxes as predicted by Madsen.

\subsection{Interactions of strangelets in matter}

Because of the present hypotetical status of strangelets (not yet discovered), it is very difficult to consider the correct and complete interactions mechanisms in matter.

The strangelet is characterised by an electrical charge and thus their interactions with electrons of atoms or molecules in the matter must exist. For high energies of the strangelet ($v > 2v_{Bohr}$), the relativistic Bethe formula for the energy loss per distance travelled can be used. Here, Bohr velocity refers to an electron in the innermost orbit of a hydrogen atom: $v_{Bohr}=c/137=2.2 \times 10^8$ cm/s.  At low energies of the incident particles the energy transferred to the recoil nuclei in accord with Lindhard theory \cite{Lindhard} is comparable with the electronic energy loss, thus they will contribute to excitations and ionizations. As a result of electronic interactions, ionizations or excitations are produced and thus scintillations are the effects. 

Of course, because the strangelets have quarks in their structure, strong interactions must exist. Possible interactions are predicted in literature - the strangelets has a high penetrability in atmosphere compared with the heavy ions, but the effects are contradictory suggested: i) in an option, their mass is reduced following collisions with atoms and molecules of the medium, whereas ii) another approach posits that the mass and charge increase following the accretion of nucleons. Wilk and Wodarczyk \cite{Wilk:1996me} interpret such a penetrability of strangelets as indication of the existence of very heavy lumps of SQM entering our atmosphere, which are then decreased in size during their consecutive collisions with air nuclei (i.e., their original mass number $A$ is reduced until $A = A_{crit}$) and finally decay be the evaporation of neutrons.  For strangelets the cross sections should be very small and hence a geometric size much smaller than typical nuclear size. In the work of Paulucci et al. \cite{Paulucci:2009zz} different possible mechanisms of interactions of strangelets with ordinary matter are discussed: abrasion, fusion, fission, excitation and de-excitation of the strangelets. It is shown that, although fusion may be important for low-energy strangelets in the interstellar medium (thus increasing the initial baryon number $A$), in the earth’s atmosphere the loss of the baryon number should be the dominant process. 

Alternatively, there exists a model based on the following hypothesis: in a typical interaction between a strangelet and the nucleus of an atmospheric atom, it is more probable for the strangelet to absorb neutrons from the colliding nucleus, and as a result the mass of the strangelet increases in every collision and it becomes more tightly bound \cite{Banerjee:2000ye, Banerjee:1998wk,Lazanu:2011zn}.

Another aspect is that in the case of strangelet, as projectile with mass and charge of certain sign, it can in principle capture particles of the opposite sign and form an "atomic system". In the case of strangelets with intrinsic negative charge, the probability to capture one or more protons is practically improbable. For the strangelets with positive intrinsic charge, the formation of one atomic system is possible. Thus, the determination of the average equilibrium charge during penetration through matter is a difficult problem. The problem was first discussed by Bohr \cite{Bohr1940,Bohr1948}, extrapolated by Pierce and Blann \cite{Pierce:1968zz} and general treatment was done by Ziegler \cite{Ziegler1980}.

In this paper we are interested only in the interaction rate of strangelets in the bulk volume of liquid noble gas used for their detection. The mechanisms by which the strangelets interact in atmosphere are not relevant for the present paper.

For the very small strangelets  Gilson and Jaffe \cite{Gilson:1993zs} show regularities as "shell effects" and islands of metastability that are persistent for a large intervals for the parameters of model. For the lower strangelets their results are in disagreement with some predictions analysed in this paper. This region of SQM is numerical investigated because it is important in the physics searches Beyond the Standard Model in the next generation of the neutrino experiments.

\section{Interactions in a large LAr neutrino detector }

For the next generation of neutrino experiments, liquid Argon (LAr) as a time projection chamber technology (TPC) is currently being developed and is very promising for neutrino detection. The DUNE detector will contain 4 modules of 10 kiloton of LAr each. For the existent prototypes, the two constructive options consists in volumes of about 700 tonnes each.  The detector will be very promising for strangelets detection.

\subsection{About scintillation yield in liquid argon and quenching phenomena}

Doke and co-workers \cite{Doke:2002oab} developed a model for the scintillation mechanisms in liquid argon. In accord with their suppositions, exist two distinct possible ways for scintillations in argon. Liquid and gaseous Ar are transparent to their own scintillation light.

Ideally, in liquid argon, the scintillation production process should be associated only with the existence of the atomic excitation mechanism. In fact, the scintillation from liquid argon has two decay time constants, corresponding to the decay from the singlet state of an excited dimer and the decay from its triplet state. Scintillation can also occur through a mechanism that starts with an ionized atom.

\begin{enumerate}
\item Ar$^*$:

Ar$^*$+Ar+Ar $\to$ Ar$_2^*$+Ar

Ar$_2^*$ $\to$ 2Ar + h$\nu$

\item Ar$^+$:

Ar$^+$+Ar $\to$ Ar$_2^+$

Ar$_2^+$ + e$^-$ $\to$ Ar$^{**}$ + Ar

Ar$^{**}$ $\to$ Ar$^*$ + heat

Ar$^*$ +Ar+Ar $\to$ Ar$_2^*$ + Ar

Ar$_2^*$ $\to$ 2Ar + h$\nu$

\end{enumerate}
where h$\nu$ denotes the ultraviolet photon and the heat corresponds to a non-radiative transition. In both processes, the excited dimer at the lowest excited level should be de-excited to the dissociative ground state by emitting a single UV photon (this hypothesis is not yet fully confirmed experimentally), because the energy gap between the lowest excitation level and the ground level is so large that there exists no decay channel such as nonradiative transitions.
The time constants of the singlet and triplet decays have been measured in all phases and their lifetimes are 7.0$\pm$1.0 ns and 1600$\pm$100 ns respectively for liquid argon. In LAr, the spectrum is dominated by an emission feature of 126.8 nm or 9.78 keV. Weak-emission features in the wavelength range from 145 to 300 nm were observed.

\begin{figure}[h]
\includegraphics[scale=0.5]{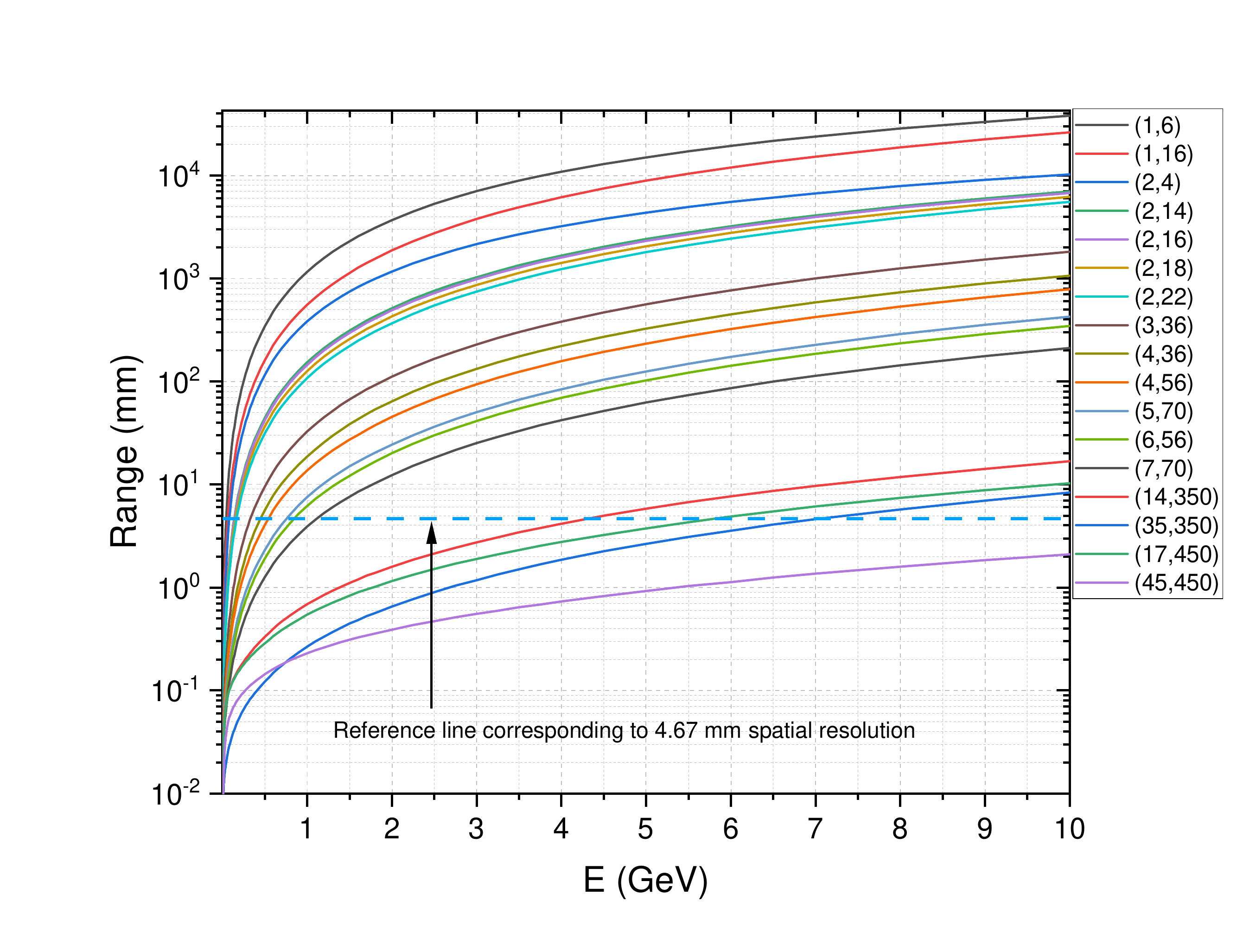}
\caption{Ranges of various Strangelets in LAr.}
\label{strangelets_range}
\end{figure}

The range of different strangelets as a function of their energies is presented in Figure \ref{strangelets_range}. 

At DUNE, as well as ProtoDUNE, the spatial resolution of the detector is 0.47 cm \cite{Abi:2020loh} and this quantity can be used as a selection to exclude candidates from the present investigation (dashed blue line in the figure).

The maximum number of scintillation photons produced for the deposited energy $E_0$ by an ionizing particle, $N_{ph}(max)$ is:

\begin{equation}
N_{ph}(max)=N_{ion}+N_{ex}=\frac{E_0}{W} \left( 1+\frac{N_{ex}}{N_{ion}} \right)=\frac{E_0}{W'}.
\end{equation} 

Considering the ionization energy $W=23.6$ eV and the mean excitation energy $W'=19.5$ eV, the total energy loss through interactions with the atomic electrons can be written as:

\begin{equation}
\begin{split}
&23.6 N_{ion}+19.5 N_{ex}=E_e, \\
&\frac{N_{ex}}{N_{ion}}=0.21.
\end{split}
\end{equation}

If the scintillation quenching and recombinations are not taken into account, one can derive the number of ionizations and excitations for any incident particle. 

In accord with Doke et. al. \cite{Doke:1988dp}, the estimated maximum number of photons emitted from liquid argon per deposited energy of 1 MeV is 5.13 × 10$^4$.

In all materials that emit fluorescence signals, these are nonlinear with respect to the energy deposited by ionizing particles, in particular for high ionizing particles. This behaviour is attributed to quenching of the primary excitation by the high ionization density along the particle track which leads to a decrease of the scintillation efficiency.
The Birks ”law” is defined as \cite{Birks:1951boa}:

\begin{equation}
\frac{dS}{dx}=\frac{A dE/dx}{A+ kB dE/dx} \to  
\begin{cases}
A dE/dx  & \text{for low value } dE/dx \\
\frac{A}{kB} = constant  & \text{for high value } dE/dx
\end{cases}
\end{equation}

\begin{figure}[h]
\includegraphics[scale=0.65]{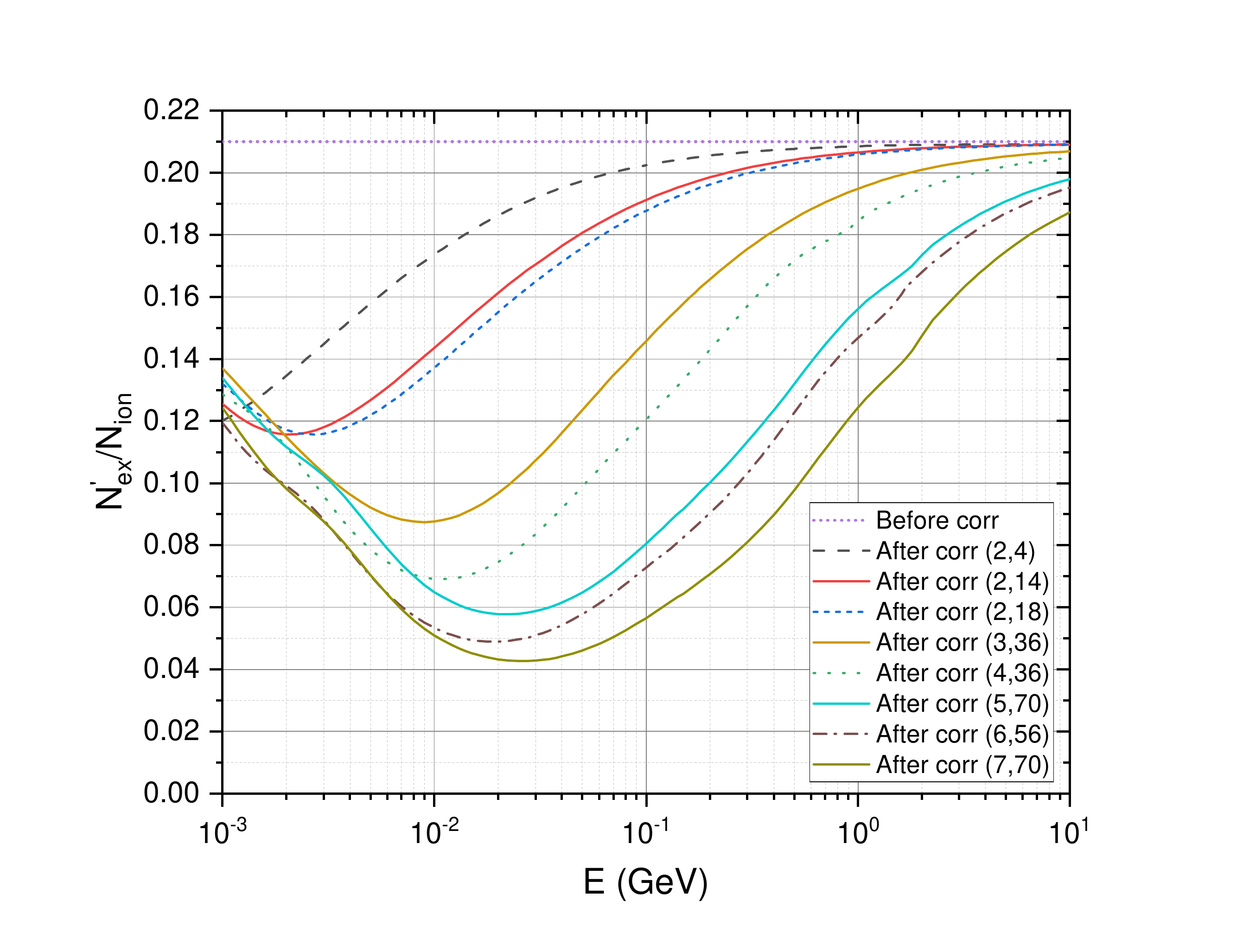}
\caption{Energy dependence of excitations vs. ionizations.}
\label{Birks}
\end{figure}

The scintillation quenching factor $f_l$ is defined by the reduction factor of scintillation intensity. As effect, the initial $N_{ex}/N_{ion}$ ratio is modified, as is indicated in Fig. \ref{Birks}, as energy dependence.
$f_l$ is defined as:

\begin{equation}
    f_l=\frac{1}{1+kB\frac{dE}{dx}},
\end{equation}
where $kB=7.4 \times 10^{-4}$ MeV$^{-1}$g cm$^{-2}$. The number of photons as a result of scintillation process after taking into account the quenching mechanism will be:
\begin{equation}
N'_{ex}=f_l N_{ex}.
\end{equation}

In a LAr detector operating as calorimeter, an electric field is applied to collect the free electrons and ions. The effects were discussed by Burdin and co-workers \cite{Burdin:2012zz}. The charge collected is then scaled appropriately to deduce the total amount of deposited energy. However, some electron–ion pairs may recombine before the ionization electron can be recorded and thus this effect reduces the recorded charge and, consequently, the deposited energy can be underestimated. The recombination effect is proportional to the ionization density, that is, the energy deposited per unit length. If $r$ is the recombination probability, a modification of Birks' formula was written \cite{Sorel:2014rka} in the form:

\begin{equation}
\begin{split}
&r=\frac{A dE/dx}{1+ B dE/dx}+C, \\
&C=1-A/B,
\end{split}
\end{equation}
where $dE/dx$ is the LET expressed in (MeV cm$^2$/g), and $A$, $B$ and $C$ are empirical parameters that depend on the noble element, density and drift electric field. For LAr, assumes:

\begin{equation}
\begin{split}
&A=0.05 \times E_{drift}^{-0.85} \mathrm{[kV/cm]}, \\
&B=0.06 \times E_{drift}^{-0.85} \mathrm{[kV/cm]}, \\
&C=1/6.  
\end{split}
\end{equation}

From the experimental point of view, the escape probability $(1-r)$ is proportional with $Q$ - the recorded charge (or "visible") and $Q_0 > Q$ is the charge produced initially.

The number of electrons produced by the ionization process, after recombinations, will be:
\begin{equation}
N'_{ion}=(1-r) N_{ion}.
\end{equation}

\begin{figure}[h]
\includegraphics[scale=0.65]{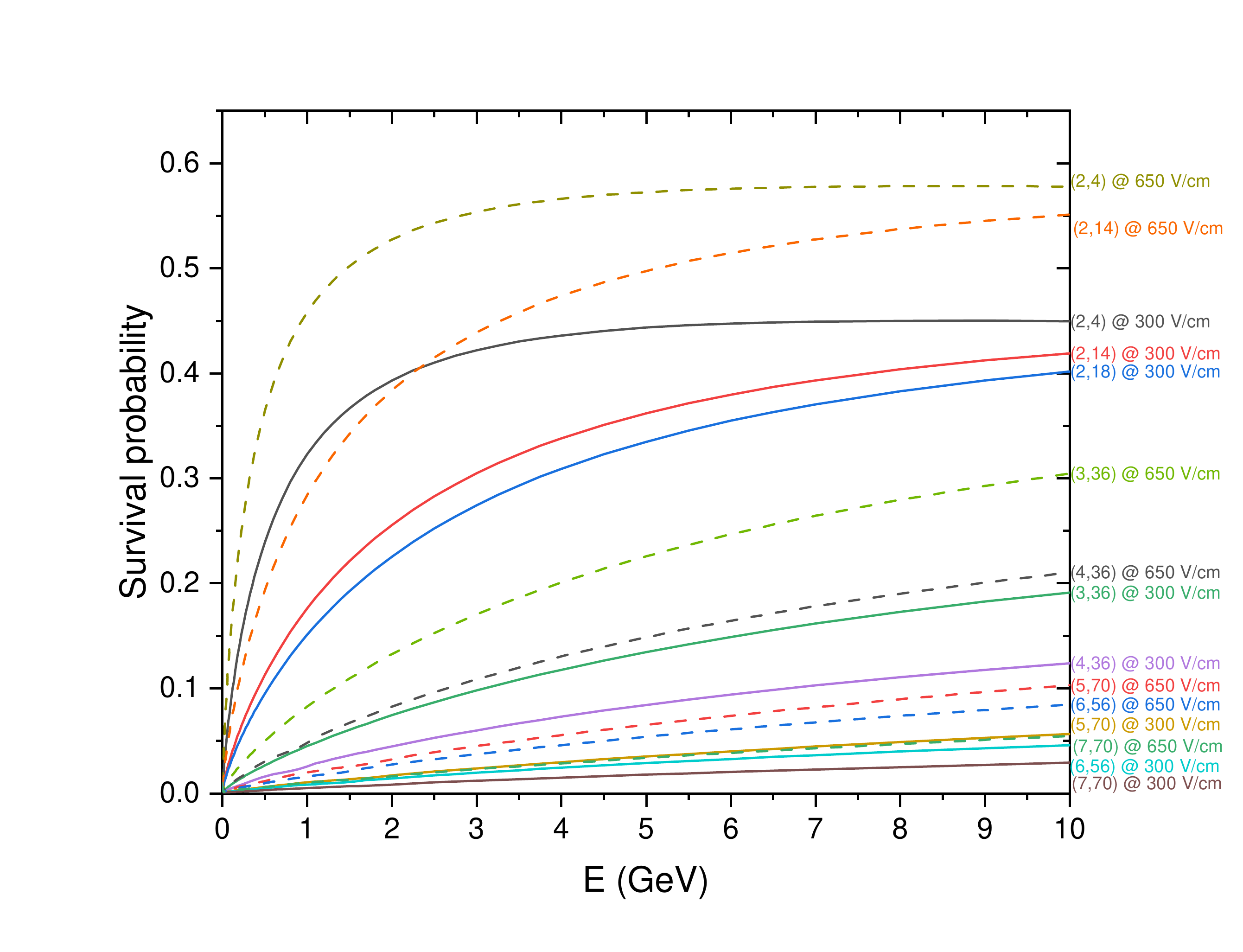}
\caption{Survival probability for two different values of the electric field (300 V/cm continuous line, 650 V/cm dashed line).} 
\label{EF}
\end{figure}

Based on the experience of the ProtoDUNE-SP the minimum intensity of the electric field (EF) required for the vertical drift (VD) is 300 V/cm; the goal is to set an EF of 500 V/cm in order to achieve the requirements of the horizontal drift (HD) and dual-phase (DP) far detectors. Experimentally the electric field for drift was increased to test up to 650 V/cm \cite{VD}. The survive probability versus strangelet's energy is represented in Fig. \ref{EF}. In this case, two different values for the drift electric field are considered: 300 V/cm and  650 V/cm respectively.

The new excitation to ionization ratio considering both the scintillation quenching and the electric field effect on the recombination processes can be expressed as:
\begin{equation}
\frac{{N'}_{ex}}{{N'}_{ion}}=\frac{N_{ex}}{N_{ion}}\frac{f_l}{(1-r)}
\end{equation}

\begin{figure}[h]
\includegraphics[scale=0.65]{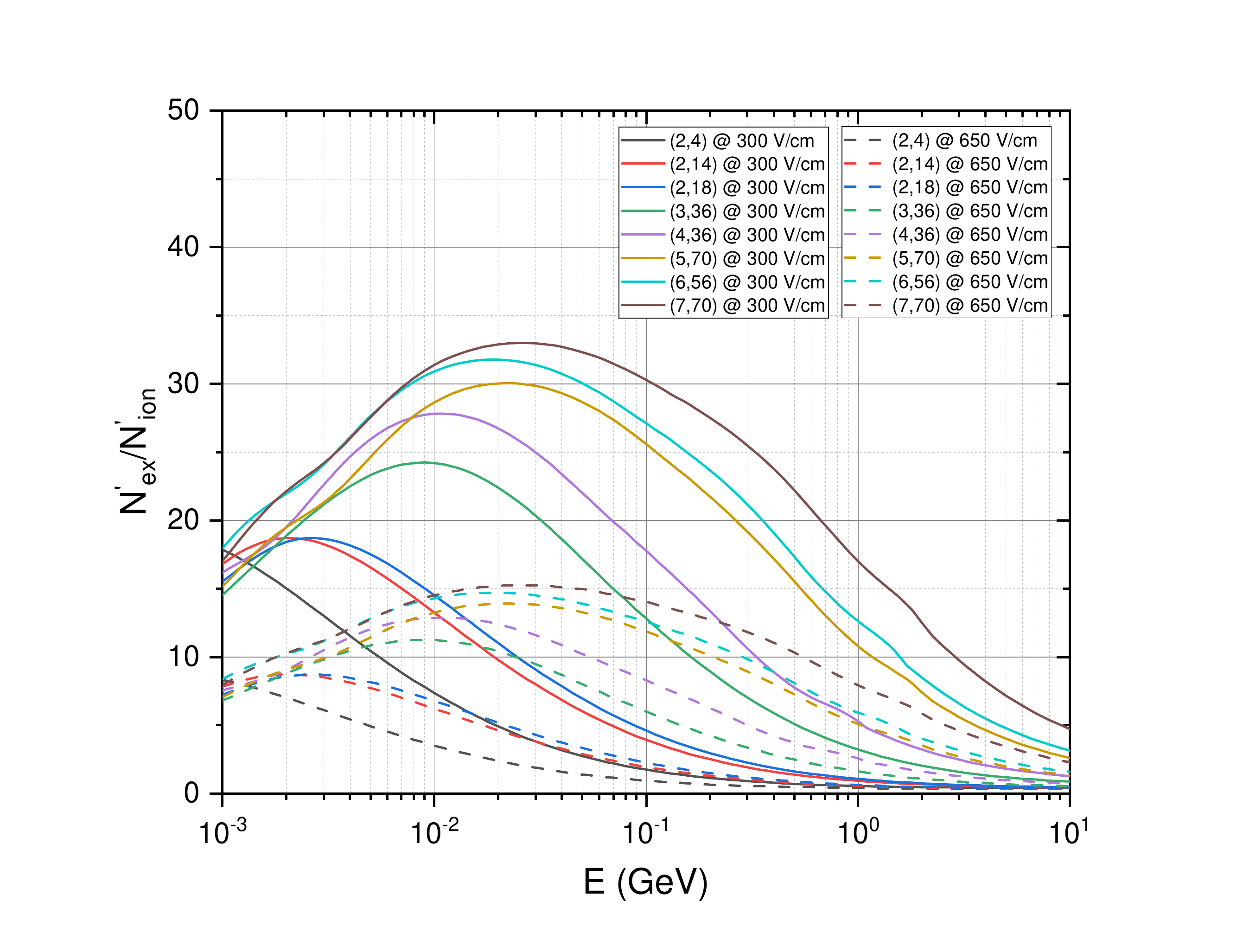}
\caption{Excitations to ionization ratio after considering both the scintillation quenching and recombinations inside two values of the electric field.}
\label{Birks+EF}
\end{figure}

Excitations to ionization ratio after considering both the scintillation quenching and recombinations are indicated in Fig. \ref{Birks+EF} for two distinct values of the drift electric field.

Except the  energy range 1-3 MeV, for the entire considered  region, the corrected ratio between the number of photons generated by the scintillation mechanism and the number of ionization electrons can be used for the discrimination between the incident projectiles with different values for ($Z$, $A$). The increase of the electric drift field considerably decreases the $N_{ex}/N_{ion}$ ratio, especially for very light strangelets if their energies are in the range of 0.1-10 GeV. In all the previous calculations of $dE/dx$ and ranges the SRIM code was used \cite{SRIM}.

\begin{figure}[h]
\includegraphics[scale=0.5]{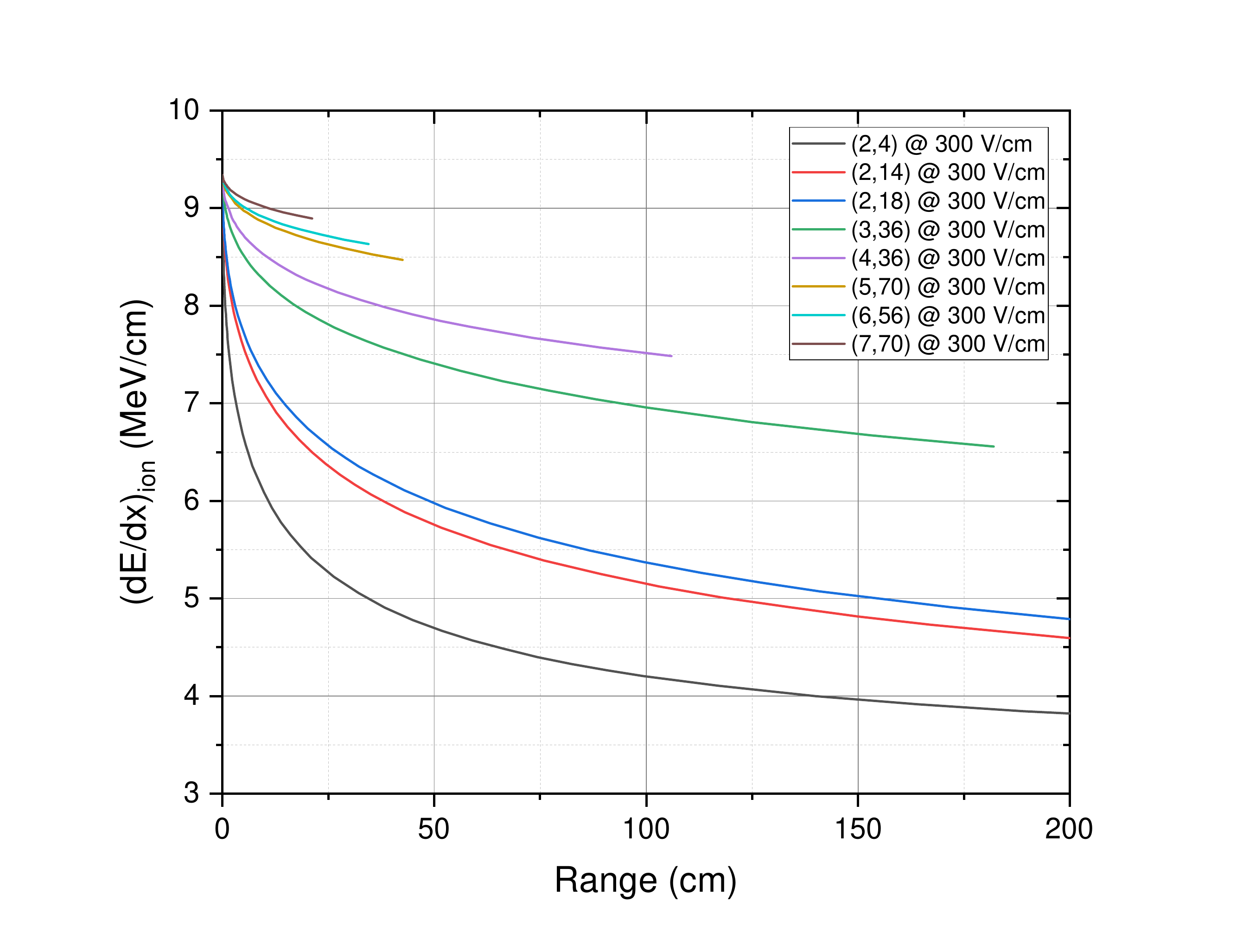}
\caption{The linear energy loss via ionizations for the considered strangelets with incident energies up to 10 GeV.}
\label{dEdx_range}
\end{figure}

When the incident strangelet (as an electrically charged particle) enters the active medium of the detector, it transfers its energy and it is partitioned between interactions with atomic electrons generating ionizations and excitations and interactions with the nuclei of the medium. Due to the kinetic energy received by the recoil nuclei, the same types of processes as the primary projectile are generated.

%\textit{In conclusion, when the incident strangelet (as an electrically charged particle) enters the detector medium, it transfers an amount of energy that is partitioned between interactions with atomic electrons generating ionizations and excitations and interactions with the nuclei of the medium. Due to the kinetic energy received by the recoil core, it partitions its energy through the same type of process as the primary projectile. Separately, we can discuss recombination and quenching processes.}
The passage of atomic particles through a target involves scattering and energy loss due to nuclear and electronic interactions. In the present discussion the contributions from energy losses in nuclear interactions was not considered because for all kinetic energies range of the projectile considered in the present paper, the nuclear cross sections are approximatively two order of magnitude below electronic cross sections. 

An essential aspect to be discussed is the ability of such a detection system to discriminate unambiguously between experimental signals from strangelets and the radioactive background from intrinsic radioactive isotopes and from cosmic background. 

The detection principle consists in measuring the signal generated by the ionization electrons and the photons coming from scintillations allowing a three-dimensional reconstruction of the event.

In Figure \ref{dEdx_range} the linear energy loss, $dE/dx$, versus the ranges of light strangelets is presented. At the same time, the value of the collected electric charge can be unequivocally correlated with the range of event, this representing an additional information in identification.

A distinct problem is the capability of these detectors to discriminate between the signals of interest and the background. In accord with Farnese \cite{Farnese:2019xgw}, ICARUS Collaboration has estimated the background value to be around $dE/dx \simeq 2.1$ MeV/cm. Thus, a clear discrimination between the considered strangelets and the background can be done.

If we also consider the scintillation signals for the considered energy range below 10 GeV, for strangelets with ($Z$, $A$) between (2,14) and (7,70), the number of photons is $> 10^5$ and can be completely discriminated from each other. In LAr detectors, the intrinsic radioactive background is dominated by the effect of the content of $^{39}$Ar in natural atmospheric argon. As a superior limit, if the purification is not realized, the measured content of radioactive $^{39}$Ar isotope will be $(8.0 \pm 0.6)\times 10^{-16}$ g/g, or $(1.01 \pm 0.08)$ Bq/kg. The undesirable events induced by $^{39}$Ar are supposed to be uniformly distributed in the active LAr volume, with an energy range up to 565 keV, the $\beta$ end-point energy, leading to very short tracks ($\sim 1$ mm) \cite{DUNE:2020txw}. The emitted electrons are a source of noise for ionizations (irrelevant in this case) and for scintillations. In order to reduce the level of $^{39}$Ar, the use of underground argon is required \cite{Church:2020env}. In the DarkSide-50 Experiment the $^{39}$Ar activity was demonstrated to be 0.73 mBq/kg, that is 1400 lower than the activity of the atmospheric argon \cite{DarkSide:2015cqb}. If we consider this case, this component of the radioactivity introduces as a maximal value, a noise around 200 photons, considerably below the signal generated by the events of interest. In this estimation we supposed a tank of 10 kt of argon.  DUNE requires an average light yield of $>20$ photoelectrons/MeV with a minimum of 0.5 photoelectron/MeV, corresponding to a photon detection efficiency (PDE) of 2.6\% and 1.3\% \cite{DUNE:2020txw}, respectively.

\section{Summary}

In this paper we investigated the possibility that light strangelets, with energies up to 10 GeV, can be directly observed as particles passing through large volumes of LAr used as the active volume of a detector.

Light strangelets were investigated, this choice being justified because experimentally there are a number of candidates as singular events reported in experiments such as AMS-01, even though their existence is theoretically controversial.

In this study the different experimental information obtained with a detector operating as a calorimeter in the TPC mode were discussed. The use of signals generated by electrons produced in the ionization process and scintillations from atomic excitations, as well as the measurements of the linear energy loss and the corresponding ranges, lead to the potential of such a detector to discriminate between the signals of strangelets after ($Z$, $A$) at kinetic energies up to 10 GeV - the limit considered in this study. It was showed that lower electric drift field is beneficial for the proposed purpose.

\section{Acknowledgment}

This work was performed with the financial support of the contract no. 02/2020, PNCDI III 2015-2020, Programme 5, Module CERN-RO. 

I.L. would like to thank UEFISCDI for partial support under project number PN-III-P1-1.1-PD-2019-0178.

\end{document}